\documentstyle[12pt]{article}
\def\fnote#1#2{\begingroup\def\thefootnote{#1}\footnote{#2}
    \addtocounter{footnote}{-1}\endgroup}
\def\email{mabe@th.phys.titech.ac.jp}
\def\sato{sato@th.phys.titech.ac.jp}
\def\IP{\relax{\rm I\kern-.18em P}}
\def\CP{\relax {\bf \rm C}{\rm I\kern-.18em P}}
\def\WCP{\relax{\bf \rm WC}{\rm I\kern-.18em P}}
\def\R{\relax{\rm I\kern-.18em R}}
\def\K3{\relax{\rm K3}}
\def\CY{\relax{\rm CY}}
\def\Pic{\relax{\rm Pic}}
\def\T{\relax{\bf \rm T}}

\begin{document}
\pagestyle{empty}
\begin{flushright}
        TIT/HEP-370
\end{flushright}
\vspace{18pt}
{\large
\begin{center}
{ \bf Calabi-Yau manifolds constructed \\
\vspace{4pt}
 by Borcea-Voisin method  
} \\ 
\vspace{8pt}

\vspace{16pt}
Mitsuko Abe \fnote{*}{\email}~ and ~Masamichi Sato\fnote{\dag}{\sato} \\

\vspace{16pt}

{\sl Department of Physics\\
Tokyo Institute of Technology \\
Oh-okayama 2-12-1, Meguro, Tokyo 152-8551, Japan}
\vspace{12pt}

{\bf ABSTRACT}
\vspace{12pt}

\begin{minipage}{4.8in}
We  construct  Calabi-Yau manifolds and  their mirrors 
from K3 surfaces.  This method was first developed by 
Borcea and Voisin. We examined their properties torically
and checked mirror symmetry  for Calabi-Yau 4-fold case.
From Borcea-Voisin 3-fold or 4-fold  
 examples, it may be possible to probe the S-duality of Seiberg
-Witten.

\end{minipage}

\end{center}
}
\vfill
\pagebreak

\pagestyle{plain}
\setcounter{page}{1}

\baselineskip=16pt

\section{Introduction}
In recent development of string duality, the mathematical properties of
the underlying manifold, on which theories are compactified, are playing  
  significant role
\cite{vafa1,vafa2,vafa3,witten0,sen,bershadsky,porrati,seiberg,fd,hori}.
In this paper,  
 Calabi-Yau manifolds and their mirrors are constructed 
 using the method developed by Borcea 
and Voisin~\cite{borcea,voisin}. 
Gross and Wilson ~\cite{gross} showed that  Calabi-Yau 
3-fold  by Borcea-Voisin 
method is a special Lagrangian three tori fibered when 
$\Pic(\K3)=U$ by using degenerate Calabi-Yau metrics. 
The  special Lagrangian fibration 
 also exist on one family of Borcea-Voisin three fold 
 with respect to non-degenarate Calabi-Yau metrics \cite{lu}. 
From Borcea-Voisin 3-fold or 4-fold  
 examples, it may be possible to probe the S-duality of Seiberg
-Witten\cite{seiberg0}\footnote{It has been pointed out that 
T-duality on the special tori causes local mirror 
transformation\cite{zaslow}, which may relate the S-duality of 
Seiberg-Witten.} by using compact manifolds.
\par 
In the next section, we give the list  
of some mirror pairs of K3 as the reflexive 
pyramids.
In  section 3, we give the list of the mirror pairs of 
Calabi-Yau 3-folds constructed by 
Borcea-Voisin method in weighted projective manifolds.
In section  4, we give the list of the mirror pairs of 
Calabi-Yau 4-folds constructed by 
Borcea-Voisin method.
Section 5, is devoted to  Discussions.
In Appendix A, we review  Borcea-Voisin method by 
using some polynomials of Calabi-Yau 
manifolds.
In Appendix B, we show a way  of the  mirror check of Calabi-Yau 4-folds.
In Appendix C, we examine properties of two  dual polyhedrons of   
 Calabi-Yau 3-fold and Calabi-Yau 4-fold constructed by Borcea-Voisin.
Especially, we present a dual polyhedron and polyhedron of 
Calabi-Yau 3-fold with Pic(K3)=U case.


\newpage
\section{K3 surface}
We start by looking at a definition of K3 surface~\cite{griffiths}.
A K3 surface is defined as a compact manifold of complex dimension
two with trivial canonical bundle such that $h^{0,1}($K3$)=0$.  
$h^{p,q}($K3$)$  denote the Hodge number of a K3 surface.

For this paper, we will consider   algebraic K3 surfaces defined by 
a set of algebraic equations in 
a $N$ dimensional complex (weighted) projective space, \CP$^N$.
The reason why we consider them  is that
weighted projective space  is  easy to describe torically. 
Such surfaces have been classified by Reid and Yonemura~\cite{yonemura}. 
The equations of K3 surface which 
we will discuss are given by table 1.
\begin{table}[h]
\[
\begin{array}{|l|l|l|} \hline
{\rm surface}&\K3\,{\rm or}\,\K3^*  &{\rm Equation~ of~ K3}\\ \hline
(1)          &\CP^3(6,4,1,1)[12]  &w^2=s^3+t^{12}+u^{12}\\ \hline
(1)^*        &\CP^3(33,22,6,5)[66]&w^2=s^3+t^{11}+tu^{12}\\ \hline
(2),(2)^*,(3)&\CP^3(21,14,6,1)[42]&w^2=s^3+t^7+u^{42}\\ \hline
(3)^*        &\CP^3(18,12,5,1)[36]&w^2=s^3+t^7u+u^{36}\\ \hline
(4)          &\CP^3(5,2,2,1)[10]  &w^2=s^5+t^5+u^{10}\\ \hline
(4)^*        &\CP^3(20,8,7,5)[40] &w^2=s^5+t^5u+u^{8}\\ \hline
(5)          &\CP^3(10,5,4,1)[20] &w^2=s^4+t^5+u^{20}\\ \hline
(5)^*        &\CP^3(15,7,6,2)[30] &w^2=s^4u+t^{5}+u^{15}\\ \hline
(6),(6)^*    &\CP^3(15,10,3,2)[24] &w^2=s^3+t^{15}+u^{10}\\ \hline
(7)          &\CP^3(12,8,3,1)[30]&w^2=s^3+t^8+u^{24}\\ \hline
(7)^*        &\CP^3(21,14,5,2)[42]&w^2=s^3+t^8 u+u^{21}\\ \hline
(8),(9)      &\CP^3(9,6,2,1)[18]  &w^2=s^3+t^9+u^{18}\\ \hline
(8)^*        &\CP^3(18,11,4,3)[36]&w^2=s^3u+t^9+u^{12}\\ \hline
(9)^*        &\CP^3(24,16,5,3)[48]&w^2=s^3+t^9u+u^{16}\\ \hline
\end{array}
\label{k30}
\]
\caption{Equation of K3 surfaces}
\end{table}
The numbers in parenthesis in the second column  denote weights of 
K3 surfaces.
\newpage 
The superscript  $\ast$  denotes the mirror
\footnote{There are some definitions of mirror symmetry 
for K3 case\cite{arnold,dolgachev,batyrev}. 
The relation between them and the extension of 
weight duality on K3 case are given in ref. \cite{kobayashi}.
Their\\ physical applications were discussed in \cite{yang}.}  
of the  corresponding  K3 surface.
 We construct Calabi-Yau 3- and 4-fold, using the
method of Borcea and Voisin~\cite{borcea,voisin}.
These manifolds have  some nice properties. 
One of them is that 
some fibers of mirror pairs are known. 
Borcea used these K3 surfaces which
 allow involution to construct 
Calabi-Yau 3-folds (the way of involution $\sigma$ is 
reviewed in Appendix A ).
\begin{table}[h]
\[
\begin{array}{|l|l|l|l|l|}\hline
{\rm surface}       &\Pic (\K3)& \rho&(r,a,\delta)
                                          &{\rm quoti.-sing.\,}\ \\ \hline
(1)     &U    &2   &(2,0,0) &A_1               \\ \hline
(1)^* &U\oplus E_8^2 &18&(18,0,0)&A_1+A_2+A_4+A_{10}\\ \hline
(2),(2)^*,(3)
      &U\oplus E_8 &10  &(10,0,0)&A_1+A_2+A_6       \\ 
\hline
(3)^* &U\oplus E_8  &10 &(10,0,0)&A_4+A_5       \\ 
\hline
(4)     &U(2)\oplus D_4&6 &(6,4,0) &5A_1              \\ 
\hline
(4)^*  &E_8\oplus T_{2,5,5}&18 &(14,4,0)&A_3+2
A_4+A_6      \\ \hline
(5)    &T_{2,5,5}&10 &(6,4,0) &A_1+2A_4     \\ \hline
(5)^*  &D_4\oplus D_8\oplus U&14 &(14,4,0)&5A_1+A_2+A_6    
  \\ \hline
(6),(6)^*   & D_4 \oplus E_6 \oplus U &12&(10,4,0)&3A_1+2A_2+A
_4     \\ \hline
(7)    &U\oplus E_6 &8 &(6,2,0) &2A_2+A_3          \\ 
\hline
(7)^* &E_8 \oplus D_4 \oplus U&14 &(14,2,0)&3A_1+A_4+A_6    
  \\ \hline
(8),(9)     &U\oplus D_4  &6 &(6,2,0) &3A_1+A_2          
\\ \hline
(8)^* &E_6 \oplus E_8 \oplus U&16 &(14,2,0)&A_1+2A_2+A_{10} 
   \\ \hline 
(9)^* & E_6\oplus E_8 \oplus U &16 &(14,2,0)&2A_2+A_4+A_7    
  \\ \hline 
\end{array}
\]
\label{k31}
\caption{K3 surface and Picard lattice. }
\end{table}
The K3 and K3$^\ast$  surfaces in table 1 satisfy the  
"reflexive pyramid" property.
This is a  sufficient condition for a pair of 
Calabi-Yau 3-folds constructed  by Borcea method  to be a mirror pair~ 
\cite{borcea}.
This condition is  stronger  than  that of K3 and K3$^*$  being 
a mirror pair.

\par
The properties  of our K3 surface are summed in table 2. 
The second column is the Picard lattice \footnote{We used the results of 
~\cite{belcastro} for Picard lattice and $\rho$.}  and the third column is  
rank of Pic(K3) denoted by $\rho$.
Note that all K3 surfaces  in table~2 are all elliptic fibered 
~\cite{belcastro}.  
The fourth column is the lattice invariants $( r, a, \delta)$. 
Nikulin  used these lattice invariant  to characterize the 
fixed part $L^\sigma$ of $\sigma$ on the K3 lattice,  
up to the lattice isomorphism \cite{nikulin1}.
 ${\bf H}_{\bf Z}^2(K3)=L=U^3 \oplus (-E_8)^2$.
 $a$ is defined as 
$ (L^\sigma)^\ast / L^\sigma\simeq ({\bf Z}/2{\bf Z})^a$. 
$(L^\sigma)^\ast$ is the dual of 
$L^\sigma$, namely 
$(L^\sigma)^\ast$= Hom ($L^\sigma,{\bf Z})$. 
 $r=rank(L^\sigma)$. 
$\delta$ denotes the 
genus of the lattice, that is,
$\delta =0$ if 
$(x^\ast)^2\in {\bf Z}$ for any 
$x^{\ast}\in$ 
$(L^\sigma)^\ast$,
otherwise $\delta =1$. 
The fifth  column is  the   quotient singularity ~\cite{yonemura}.

\newpage  
\section{Calabi-Yau 3-fold}

\subsection{K3 fibered Calabi-Yau 3-fold}
Let us now consider   Calabi-Yau manifolds with base, 
\CP$^1$=\CP$^1$(1,k)
\footnote{  The equivalence for  
type IIA string on CY$_3$ with  $k \geq 2$ 
dual to Heterotic string on K3 $\times$ T$^2$ 
 is not clear  yet.  
 However, some extension of the duality   
 to $k \geq 2 $ and $u_1=1$ may be possible.  
For example,
 $\CY_3=\CP^4(1,k,(k+1)(1,4,6))[12(k+1)]$ with 
$\K3=\CP^3(1,1,4,6)[12]$ fiber   
relate to the terminal A-chain with shrinking E$_8$ instantons 
\cite{candelas} by  exchanging  of base under the  elliptic 
fibration, that is,   F$_0$ blown-up  and F$_2$ 
blown-up~\cite{mabe}. They are the same manifolds with 
double K3 fibrations.} 
 and fiber, $\K3=\CP^3(u_1, u_2, u_3, u_4)[d]$~\cite{yau}.
They are represented as hypersurface in weighted projective 4-space, \CP$^4$, 
\begin{equation}
\CY_3= \CP^4(u_1, k u_1, (k+1)u_2, (k+1)u_3, (k+1)u_4))[(k+1)d],
\end {equation}
where $ d= \sum_{i=1}^4 u_i$.
\par
There are some mirror pairs of Calabi-Yau 3-folds, which have mirror pair
of K3 surfaces as fiber.
See, for instance, the self-mirror Calabi-Yau 3-folds
(2), (6) in table~5
 \footnote {These Calabi-Yau 3-folds 
were already investigated and listed in 
\cite{avram}.}.  
\noindent
These manifolds have self-mirror K3 surfaces 
$\CP^3(21,14,6,1)[42]$
and $\CP^3(15,10,3,2)[30]$ as K3 fiber.
Another example of a mirror pair is  (4)  and (4)$^\ast$, which 
 can be obtained by Borcea-Voisin 
method\footnote{All  Calabi-Yau 3-folds given in ref.~\cite{yau} 
are smooth.
However, by picking up appropriate terms, we obtain Calabi-Yau 3-fold which 
we are treating.}.
\par

\begin{table} 
\[
\begin{array}{|l|l|l|l|l|}\hline
{\rm surface} &{\rm (\CP^1~ base,~ \K3~ fiber)}                &\CY  _3      &h^{1,1},\,h^{1,2}&\chi\\ 
\hline
(2)&(\CP^1(1,1),\CP^3(21,14,6,1)[42])&\CP^4(21,21,2(14,6,1))[84]&35,\,35&0\\ 
\hline
(4)&(\CP^1(1,2),\CP^3(5,2, 2, 1)[10])&\CP^4( 5,10,3(2,2, 1))[30]&15,\,39&-48
\\ \hline
(4^*)&(\CP^1(1,2),\CP^3(20,8,7, 5)[40])&\CP^4(20,40,3(8,7, 5))[120]&39,\,15&48
\\ \hline
(6)&(\CP^1(1,1),\CP^3(15,10,3,2)[30])&\CP^4(15,15,2(10,3,2))[60]&27,\,27&0\\
\hline
\end{array}
\label{cy3}
\]
\caption{Calabi-Yau 3-fold 1.}
\end{table}


The word ''self-mirror manifold'' 
contains a  deformed   manifold from  the strict self-mirror   
one  whose faces are lattice equivalent to vertices.   
 Therefore, in case (2) and case (6), 
 their faces  are not lattice equivalent to vertices.

\subsection{Borcea-Voisin construction}
We use the K3 surfaces listed in table~2.
In the above equation, $(-1)$ denotes involution acting on $\T^2$ 
and $\sigma$ on K3 surface. 
For K3 surface, the involution  
 changes the sign of one of the coordinates describing the torus in 
elliptic fiber. 
\begin{equation}
\CY_3={\T^2 \times \K3 \over (-1) \times \sigma}, ~  
\CY_3^\ast ={\T^2 \times \K3^\ast \over (-1) \times \check \sigma}.
\end{equation}
For the details of this  construction, see ref.~\cite{borcea} and 
appendix A\footnote{
In ref.~\cite{schimmrigk1,schimmrigk2}, Calabi-Yau manifolds are
constructed by  the  extended way of the Borcea method using K3 surfaces 
listed in~\cite{kondo}.}.
Because of the condition gcd$(u_0,v_0)=1$ (see appendix A), we obtain
only  three mirror pairs as the hypersurfaces in  CP$^4$ 
with weight representations.
The results of the construction of mirror pairs of  Calabi-Yau 3-folds  
are given by table 4. However, it is possible to obtain other mirror pairs 
of Calabi-Yau 3-folds by using all K3 surfaces in table 1 and their 
toric data.  This is because  one side of Calabi-Yau 3-folds can be 
represented by the hypersurface representation in terms of weight 
by using either  T$^2$=CP$^2$(1,1,2)[4] or T$^2$=CP$^2$(1,2,3)[6] at least.  
For example, toric data of a Calabi-Yau 3-fold  
 contructed from  K3=CP$^3$(1,1,4,6)[12]  with Pic(K3)=U  
in case (1) of table 1 can be  obtained (  in  Appendix C ). 
This is  the case with special three tori fibered \cite{gross}.

\begin{table}[h]
\[
\begin{array}{|l|l|l|l|} \hline
{\rm surfaces} &\T^2~{\rm fiber} &\K3~{\rm fiber}&CY_3\\ \hline
(2)&\CP^2(2,1,1)[4]&\CP^3(1,6,14,21)[42]&\CP^4(21,21,28,12,2)[84]
\\ \hline
(4)&\CP^2(3,2,1)[6]&\CP^3(5,2,2,1)[10]  &\CP^4(5,10,6,6,3)[30]
\\ \hline
(4^*)&\CP^2(3,2,1)[6]&\CP^3(20,8,7,5)[40]&\CP^4(20,40,24,21,15)[120]
\\ \hline
(6)&\CP^2(2,1,1)[4]&\CP^3(15,10,3,2)[30]&\CP^4(15,15,20,6,4)[60]
\\ \hline
\end{array}
\label{cy4}
\]
\caption{Calabi-Yau 3-fold 2.}
\end{table}
\noindent
Equations of these manifolds are listed in  table 5.
\begin{table}[h]
\[
\begin{array}{|l|l|l|l|l|} \hline
{\rm surface} &{\rm Equation\, of\, \CY_3}&h^{1,1},h^{1,2}
&\chi & {\rm quoti.~ sing.}
\\ \hline
(2)&x^4+y^4=s^3+t^7+u^{42}& 35,~35&0&4A_2+4A_6+A_1\\ \hline
(4)&x^3+y^6=s^5+t^5+u^{10}&15,~39&-48&5A_2+5A_1+3A_4\\ \hline
(4^*)&x^3+y^6=s^5+t^5u+u^8&39,~15&48&3A_2+A_1+A_6\\ \hline
(6)&x^4+y^4=s^3+t^{15}+u^{10}&27,~27&0&A_1+4A_2+4A_4\\ \hline
\end{array}
\label{curve}
\]
\caption{Equation of CY$_3$}
\end{table}
The Hodge numbers given by \cite{borcea} is represented 
by $r$, $f$ or $g$, whose values coincide with the  calculation 
using  the adjunction formula, 
\begin{equation}
h^{1,1}(\CY_3)=1+r+4f,~
h^{1,2}(\CY_3)=1+(20-r)+4g,
\end{equation}
where, 
\begin{equation}
f=(r -a)/2+1 ~{\rm and}~ g=(20- r-a )/ 2+1.
\end{equation}
\noindent
By mirror transformation,  
 the  values of $r$ become  $20-r$ and the values of $f$ and $g$ 
are exchanged.
\par 
In Calabi-Yau 3-folds constructed by Borcea-Voisin, 
there are  two or four fixed points generated by the 
involution on $\T^2$ or on S$_1$~
\cite{vafa3,gross}. 
S$_1$ denotes one-dimensional sphere.
If  Calabi-Yau 3-folds have 
the elliptic fibration due to the one side of  T$^2$ part of the 
direct product  only, then
 the singularities of elliptic fiber are 
$SO(8)^f$ for Calabi-Yau 3-fold or $SO(8)^g$ for Calabi-Yau 
3-fold$^\ast$  in addition to $ SU(3)$ \cite{vafa3,gross}. 
We conclude that 
though the singularity of the elliptic fibration  of  K3 fiber
 might appear due to the quotient singularity of Calabi-Yau 3-folds or 
due to the tensor multiplets when base under K3 fibration being large 
\footnote{  The number of the tensor multiplets in D=6 and N=1 are 
given by $r-2$ ( or $20-r -2$) for Calabi-Yau 3-fold 
( or the mirror of Calabi-Yau 3-fold)  \cite{vafa3} when $T^2$ 
fiber from the direct product part of T$^2$ is used in the compactification.  
They lead to   the same numbers of the 
U(1) vector multiplets  for D=4  and N=2 case from type IIA side, 
which might enhance. }, 
they  will not appear  
 as the  singularity of the elliptic fibration of Calabi-Yau 3-fold.
\newpage
(We  examine the possibility of existing  a 
 elliptic fibration coming from   K3 fiber side  in the Appendix C. ) 
\footnote{If Calabi-Yau 3-fold by Borcea-Voisin is 
double K3 fibered, then 
case (4) may   relate to  the  Heterotic  and 
type IIA string duality  
in C-chain  whose  gauge symmetries is  
$U(1)^2 \times D_5 $  with $H^{1,1}=15$ and $H^{1,2}=39$  \cite{candelas}.}.

\newpage
\section{Calabi-Yau 4-fold}
In this section, we  consider  
three types of Calabi-Yau 4-folds.
\newline
\null
\noindent

\noindent
The first basic examples are composed with two K3 surfaces with 
involution given by \cite{borcea}. 
\begin{equation}
\CY_4={\K3 \times \widetilde {\K3} \over \sigma\times \sigma^\prime},~
\CY_4^\ast={\K3^\ast \times \widetilde {\K3}^\ast  \over
\sigma\times \sigma^\prime},
\end{equation}
The involution $\sigma$  acts  on K3 and $\sigma'$
 on K3$^*$.
Note that the condition 
\linebreak
gcd$(u_0, v_0)=1$ restricts the choice of K3
pairs.
Table 6 contains all mirror pairs  with weight representations 
  constructed from K3 surfaces listed in table~2. 
We can  prove that these pairs  are mirror  by using their  
polyhedra derived from  the weight representation  in Appendix B.
  
\begin{table}[h]
\[
\begin{array}{|l|l|l|l|}\hline
{\rm surface} &{\rm K3~ fiber}&{\rm \widetilde{K3~ fiber}}&{\rm CY_4}\\ \hline
(1)&\CP^3(5,2,2,1)[10]&\CP^3(6,4,1,1)[12]&\CP^5(20,5,5,12,12,6)[60]
\\ \hline
(1)^*&\CP^3(20,8,7,5)[40]&\CP^3(33,22,6,5)[66]&\CP^5(440,120,100,264,231,165)[
1320]
\\ \hline
(2)&\CP^3(5,2,2,1)[10]&\CP^3(21,14,6,1)[42]&\CP^5(70,30,5,42,42,21)[210]
\\ \hline
(2)^*,(3)&\CP^3(20,8,7,5)[40]&\CP^3(21,14,6,1)[42]&\CP^5(280,120,20,168,147,10
5)[840]
\\ \hline
(3)^*&\CP^3(5,2,2,1)[10]&\CP^3(18,12,5,1)[36]&\CP^5(60,25,5,36,36,18)[180]
\\ \hline
(7)&\CP^3(5,2,2,1)[10]&\CP^3(12,8,3,1)[24]&\CP^5(40,15,5,24,24,12)[180]
\\ \hline
(7)^*&\CP^3(20,8,7,5)[40]&\CP^3(21,14,5,2)[42]&\CP^5(280,100,40,168,147,105)[8
40]
\\ \hline
(8)^*&\CP^3(5,2,2,1)[10]&\CP^3(18,11,4,3)[36]&\CP^5(55,20,15,36,36,18)[180]
\\ \hline
(8),(9)&\CP^3(20,8,7,5)[40]&\CP^3(9,6,2,1)[18]&\CP^5(120,40,20,72,63,45)[360]
\\ \hline
(9)^*&\CP^3(5,2,2,1)[10]&\CP^3(24,16,5,3)[48]&\CP^5(80,25,15,48,48,24)[240]
\\ \hline
\end{array}
\label{cy43}
\]
\caption{ Calabi-Yau 4-fold}
\end{table}


The Hodge numbers of Calabi-Yau 4-folds   given  by 
\cite{borcea} are
\begin{eqnarray}
 &h^{1,1}=&r_1+r_2+f_1f_2, \nonumber \\ 
 & h^{3,1}=&(20-r_1)+(20-r_2)+g_1g_2, \nonumber \\ 
 & h^{2,1}=&f_1g_2+f_2g_1, \nonumber \\ 
 & h^{2,2}=&2[102+(r_1-10)\cdot(r_2-10)+f_1f_2+g_1 g_2], \nonumber \\  
\end{eqnarray}
where $f_i$ and $g_i$ are given by eq. (13). 
The suffix $i$ specifies K3 or K3$^\ast$ in eq.(14).
The Euler number of  Calabi-Yau 4-fold is       
\begin{equation}
\chi=4+2(h^{1,1}+h^{1,3})+h^{2,2}-4h^{1,2}.
\end{equation}

\noindent
Equations  and the Euler number  are given by table 7.
\begin{table}[h]
\[
\begin{array}{|l|l|l|l|l|l|l|l|l|l|}\hline
 {\rm surface}&{\rm Equation\, of\, \CY}_4 & h^{1,1}&h^{2,1}&h^{3,1}&h^{2,2}&\chi\\ \hline
(1)&y^3+z^{12}+w^{12}+s^5+t^5+u^{10}=0&12&32&92&396&480\\ \hline
(1)^*&y^3+z^{11}+zw^{12}+s^5+t^5u+u^8=0&92&32&12&396&480\\ \hline
(2)&y^3+z^7+w^{42}+s^5+t^5+u^{10}=0&28&48&60&300&464\\ \hline
(2)^*,(3)&y^3+z^7+w^{42}+s^5+t^5u+u^8=0&60&48&28&300&464\\ \hline
(3)^*&y^3+z^7w+w^{36}+s^5+t^5+u^{10}=0&28&48&60&300&464\\ \hline
(7)&y^3+z^8 +w^{24}+s^5+t^5+u^{10}=0&14&44&102&420&480\\ \hline
(7)^*&y^3+z^8w+w^{21}+s^5+t^5u+u^8=0&102&44&14&420&480\\ \hline
(8)^*&y^3w+z^9 +w^{12}+s^5+t^5+u^{10}=0&34&48&38&236&192\\ \hline
(8),(9)&y^3+z^9+w^{18}+s^5+t^5u+u^8=0&38&48&34&236&192\\ \hline
(9)^*&y^3+z^9w +w^{16}+s^5+t^5+u^{10}=0&34&48&38&236&192\\ \hline
\end{array}
\label{hodge42}
\]
\caption{Equation of \CY$_4$}
\end{table}

When both K3 surfaces are elliptically fibered, the dual theory 
of  F-theory compactified on Calabi-Yau 4-fold will 
be  type I$^\prime$ theory compactified on $T^2 \times K3/Z_2$.
The  enhanced gauge symmetries will be the related to the 
 singular elliptic fibers or 
the quotient singularities of a Calabi-Yau 4-fold  relating to  
 K3 singularities.

\vspace{1cm}
The second type,

\begin{equation}
\CY_4={\T^2 \times \CY^3  \over   (-1)\times \sigma },
\label{eqcy4}
\end{equation}
 can be obtained from Borcea-Voisin method
(here, Calabi-Yau 3-fold has an involution)
~\cite{borcea,schimmrigk1,schimmrigk2}.
 We can construct the following  example. 
By using  
 \CY$_3$=\CP$^4$(1,42,258,602,903)[1806] and 
\T$_2$=CP$^2$(1,1,2)[4], 
 we obtain 

\CY$_4$=\CP$^5$(2,84,516,903,903,1204)[3612]
\footnote{It is also in the list of Calabi-Yau 4-fold  \cite{avram}.}.
  Hodge numbers are  
\begin{equation}
h^{1,1}=h^{3,1}=500,~  h^{1,2 }=0~ {\rm and}~\chi=6048.  
\end{equation}

In this case, Calabi-Yau 3-fold is
K3 fibered Weierstrass type one such as 
Calabi-Yau 3-fold and K3 fiber have the same \T$^2$ fiber. 
\K3 fiber is  \CP$^3$(1,6,14,21)[42] with $E_8$ type elliptic singularity.
It may be  self-mirror   though their faces 
and vertices are not SL(5,{\bf Z}) equivalent 
\footnote{ For self-mirror families,  
vertices and  faces are not SL(5,{\bf Z}) 
equivalent and they are only  GL(5,{\bf Z}) equivalent. This 
condition may be a necessary one for  self-mirror cases.}. 
It has  $h^{1,1}=h^{2,1}=251$ and $k=42$ in eq.(1). 
CY 3-fold satisfies the condition of being 
reflexsive pyramid which is extended to the higher dimensions.
$k=42$ is the only case 
when $( k+1)903 $ is coprime with 2  or  3 
among $k=\{2,3,7,41,42 \}$ in the list of  \cite{yau}.

\null
For the third type,
the  manifold may  be obtained by the extending
Borcea-Voisin method~,\\
\begin{equation}
\CY_4=\frac{\T^2\times{\frac {\T^2 \times \K3 }{ (-1)\times \sigma } }}
{ (-1)\times \sigma }.
\label{eqcy42}
\end{equation}
Holonomy group of covering space of this manifold is $SU(2)$.
It  would be interesting to consider the application of this manifold to 
verify string duality. 
\newpage
\section{Discussion}

Many Calabi-Yau manifolds were constructed by Borcea-Voisin method. 
Furthermore, many mirror pairs are possible without weight 
representations  in addition to our list 
with weight representations\footnote{
The reason why we made mirror pairs of Calabi-Yau manifolds with 
weight representations  is that we wanted to examine their properties  
torically.   }.
\par  
These manifolds have some nice properties.
 For example, in  Calabi-Yau 3-fold case :
\newline
1)
Mirror pair can be constructed easily,
\newline
2) Mirror pair manifolds have the same \T$^2$ fiber,  
\newline
3) K3 fibers of mirror pair manifolds have the same base $\T^2/(-1)$, 
\newline
4) The number of the rational curves are known. Therefore,  the 
superpotentials are obtained in the exact forms 
\cite{voisin} in Type IIA and Type IIB side. 
The  correspondence of type IIA and type IIB is apparent.

It should  be possible to use these  CY4-folds  to see the
relation between F-theory  on Calbi-Yau 4-folds and type I$^\prime$ theory  
on Calbi-Yau 3-folds or $\T^2 \times \K3/Z_2$.  
For example, Sen~\cite{sen} used  Calabi-Yau 3-folds 
constructed by Borcea-Voisin 
method, 
($\CY_3=(\T^2,{\rm F}_m),\,m=0,\,1,\,4$) and showed the relation among 
F-theory compactified on Calabi-Yau 3-fold and type I' theory 
on K3 surface.

Studying the duality of supersymmetric field theory with brane will  
 require the clarification of  a relation among T-duality, 
mirror symmetry and Fourier-Mukai 
transformation~\cite{hori,zaslow,mukai,morris}. 
The manifolds constructed in the present paper will be applied to 
investigate  this relation in future work.

\newpage
\null
\vspace{16pt}

\noindent
{\Large{\bf Acknowledgement}}
\vspace{16pt}

\noindent
We would like to greatly thank M. Kobayashi, 
 S. Kondo, S. Hosono and Sarha-Marie Belcastro 
for comprehensive guidance of 
mathematics. 
We also owe special thanks to 
N. Sakai for useful guidance for physics and 
 K. Mohri for fruitful discussion and for the idea on    
how to check mirror when all the weights are not equal to one.
\newpage
\appendix{{\Large \bf Appendix}}
\section{Borcea-Voisin method}
We start with the manifolds with 
 weights, $u=(u_0,u_1,\cdot\cdot,u_n)$ 
and $v=(v_0,v_1,\cdot\cdot,v_m)$, 
where, 
$
u_0=\Sigma_{i=1}^n u_i\;{\rm and}\; v_0=\Sigma_{i=1}^m v_i.
$
 We assume the following form to the equations which 
describes these manifolds, 
\begin{equation}
x_0^2=f(x_1,x_2,\ldots,x_n)\; {\rm and}\; y_0^2=f(y_1,y_2,\ldots,y_m).
\end{equation}
Then the Calabi-Yau hypersurfaces $X^{(u)}_n$ and $Y^{(v)}_m$ of degree
$2u_0$ and $2v_0$, respectively are defined, 
\begin{equation}
\CP(u)=\CP(u_0,u_1,\ldots,u_n)\; {\rm and}\; \CP(v)=\CP(v_0,v_1,\ldots,v_m).
\end{equation}
Furthermore, assume that gcd$(u_0,v_0)=1$  to obtain the rational map
\begin{equation}
\CP(u)\times \CP(v)=\CP(v_0u_1,\ldots,v_0u_n,u_0v_1,\ldots,u_0v_m),
\end{equation}
defined by
\begin{equation}
(x_0,\ldots x_n)\times (y_0,\ldots y_m)
\rightarrow
(x_1y_0^{u_1/u_0},\ldots,x_ny_0^{u_n/u_0},y_1x_0^{v_1/v_0},
\ldots,y_mx_0^{v_m/v_0})
\end{equation}
where  all the fractional powers use the same determination for
$y_0^{1/u_0}$ and $x_0^{1/v_0}$ respectively.
This is a Calabi-Yau hypersurface $X_{n+m}^{u\times v}$ of total 
degree $2u_0v_0$.

We used two types of  tori for the construction of $CY_3$,
\begin{equation}
\CP^2(1,1,2)[4]~:~y_1^2=y_2^4+y_3^4,
\end{equation}
{\rm and}
\begin{equation}
\CP^2(1,2,3)[6]~:~y_1^2=y_2^3+y_3^6.
\end{equation}

\newpage
\section{The mirror check of Calabi-Yau 4-folds}
Here, we will show a simple way
\footnote{We owe this way to Mohri.} to  check mirror 
 using  a following pair of Calabi-Yau 4-folds.     
 We will use (2) and $(2)^*$ in the table 8 as an example. 
All of their weight are not equal one.
\newline
$(2)~\CP^5(70,30,5,42,42,21)[210] ~~
(2)^*~ \CP^5(280,120,20,168,147,105)[840]$.
\newline
The  coordinate  transformation  of the polyhedra\cite{batyrev} in 
the  lattice is as follows~:
\newline
$(x_1, \cdots,x_6)~ \in~ {\bf Z}^6 ~ 
\rightarrow~ (m_1, \cdots,m_5)~ \in {\bf Z}^5$.
\par
For case (2), the basic equation which they must satisfy  is 
\begin{eqnarray}
&5 (14x_1+6x_2+x_3)+21(2x_4+2x_5+x_6)
\nonumber \\
& \equiv 5 \times (-21 m_3) + 21 \times 5 m_3=0.
\end{eqnarray}
A solution is given by 
\begin{eqnarray}
&x_1=m_1,~x_2=m_2,~x_3=-14m_1-6m_2-21m_3&,
 \nonumber\\ 
&x_4=m_4,~x_5=m_5,~x_6=5m_3-2m_4-2m_5&, \forall x_i \geq -1.
\end{eqnarray}
Some similar deformations  are  possible for  $(2)^*$, 
\begin{eqnarray}
&x_1=m_1,~x_2=m_2,~x_3=-14m_1-6m_2-21m_3,~\forall x_i \geq -1&
 \nonumber\\ 
&x_4=6m_3-7m_4+2m_5,~x_5=-4m_3+8m_4 -3m_5,~x_6=m_5,& .
\end{eqnarray}
\noindent
This comes from the relation between weights and  degrees.

\par
Using   a program of  analyzing polytopes 
and polyhedra, the faces and the vertices of polyhedra  
are obtained.
For example, the faces of (2) and the vertices of  $(2)^*$ are given by  
\begin{eqnarray}
&\vec f_1=(14,6,21,0,0),&~~\vec v_1=(-1,-1,-1,-1,-1), \nonumber
\\   
&\vec f_2=(-1,0,0,0,0),&~~\vec v_2=(2,-1,-1,-1,-1), \nonumber
\\
&\vec f_3=(0,-1,0,0,0),&~~\vec v_3=(-1,6,-1,-1,-1), \nonumber
\\
&\vec f_4=(0,0,0,-1,0),&~~\vec v_4=(-1,-1,1,1,0), \nonumber
\\
&\vec f_5=(0,0,0,0,-1),&~~\vec v_5=(-1,-1,1,0,-1), \nonumber
\\
&\vec f_6=(0,0,-5,2,2),&~~\vec v_6=(-1,-1,1,3,7). 
\end{eqnarray}
\noindent
They  are lattice isomorphic, i.e., 
${}^\exists  {\bf A} \in SL(5, {\bf Z}),~ {\bf A} \vec f_i = \vec v_i$ 
\newline  for ~ 
$ i=1 \cdots 6$.

\[
A=\left(
\begin{array}{c}
-2,~~1,~~1,~~1,~~1\\
~~1,-6,~~1,~~1,~~1\\
~~1,~~1,-1,-1,-1\\
~~1,~~1,-1,-1,~~0\\
~~1,~~1,-1,~~0,~~1\\
\end{array}
\right)
\]
\noindent
Thus, (2) and $(2)^\ast$ are a mirror pair.

\newpage
\section{ 
 Dual Polyhedra of Calabi-Yau 3-fold 
in Borcea-Voison construction}

There are some works about   singularities in  
 algebraic manifolds including K3 surfaces 
 \cite{kodaira,shioda,kondo,saito,milnor,ebeling1}.  
 It is difficult  to  know  all   elliptic fibrations 
  and all degeneration which Calabi-Yau manifolds have 
  even if they are K3 surfaces.  
 Some ways of obtaining them in terms of toric varieties 
are proposed \cite{candelas,avram}
\footnote{
For K3 cases, Belcastro  gave some more   improved 
    ways  of  calculating  Pic (K3),  finding  
elliptic fibrations \cite{belcastro}.
Her ways  are    composite ones of using  toric varieties  and  
 some  theorems of algebraic geometry. 
 Her way to  calculate  the intersection number of two 
curves  is to look at the graph associated to the incidence 
matrix associated to the desingularization of the polytope. 
If the vertices representing the curve have an edge between them, 
then thier intersection multiplicity is the multiplicity of the 
edge (otherwise 0).   
She gave the list of Pic (K3) and the elliptic fibers 
for 95  K3 surfaces and  
types of generations \cite{belcastro}.}.   
 They use a dual polyhedron of Calabi-Yau manifolds to find fibrations 
and    their singularities.  
For example, in finding singularities,     
  top points  and the  bottom points in the edge 
of the dual polyhedron of K3  
denote the extended Dynkin diagrams of the  singularities.
Their method is simple and  visible. 
However, there are some ambiguities because they have  
the lattice equivalence  and  some vanishing points. 
Therefore, we think that the sufficient condition to identify 
the final step corresponding to the boundary of Kahler cone
 may be useful in using  their method. 
The boundary of the Kahler cone is where we can find 
all elliptic fiberations and their degenerations 
which Calabi-Yau manifolds have. It corresponds to the most smooth 
Calabi-Yau manifold. 
Vinberg gave an algorism to obtain the boundary of the 
Kahler cones  of K3. He  derived   the sufficient condition to  
identify the last stage of the 
algorism for the  signature (1,n) case of the 
Picard lattice \cite{vinberg}. 
 If we can translate this condition in terms of  dual polyhedron,  
we can apply this condition to the method of  refs in 
 \cite{candelas,avram}   for these 
 cases.  
From Picard lattice, it may  be possible to  obtain 
 the boundary of the Kahler cone by the following method
\footnote{
Kahler cone  can be obtained 
as the fundamental region of the Weyl transformation  of the 
Picard lattice for K3 case.}. 
The dual graph of Kahler cone ( secondary polytope )
 is obtained by 
 the triangulation of dual polyhedra \footnote
{There are  some  softwares  to get secondary polytopes. 
 Unfortunately, even for K3 cases,
if $\rho \geq 8$, then  they will not  
work unless one uses some symmetries. }.
After choosing the case which has the most triangulations 
which corresponds to the most smooth K3, we can identify it 
 as the dual of the boundary of the Kahler cone. 
If we can get the dual graph of the secondary polytope 
then we will be able to see the boundary of the Kahler cone  
  constructed by CP$^1$.  
 If we can make the  Gram matrix  
 whose element is represented  by the intersection number of 
each CP$^1$ in the boundary of the Kahler cone then 
it contains all the elliptic curves 
which are resoluted and forming   the extended Dynkin diagrams
\footnote{It may be possible to link the intersection number 
of the singularities and the secondary polytope by using  
\cite{zharkov}.}.  

We applied the methods of the references of    
\cite{candelas,avram} to investigate fibers in 
 Calabi-Yau 3-folds in table 5. 
Their method is summarized as below.
We will follow the notation of the ref. of  \cite{candelas}.  The upper 
preffix  in $ \nabla$ denotes the dimension of 
the lattice of the polyhedron or 
dual polyhedron.
\newline
\noindent
$\{(x_1,x_2,x_3,x_4)\} 
 \in {}^4\nabla $  form  integral points in   
the four-dimensional  dual polyhedron 
of Calabi-Yau 3-fold up to the points in codimension one face.
 \newline
\noindent
$\{(x_1,x_2)\}\in  {}^2\nabla \subset {}^4 \nabla   $ represent  
 the integral points in the dual polyhedron of a base under the 
elliptic fibration of Calabi-Yau 3-folds. 
\newline
\noindent
 $\{(x_2,x_3,x_4)\} \mid_{x_1=0}
 \in {}^3\nabla ~ \subset {}^4\nabla $
denote the integral points in 
 the   three-dimensional  dual polyhedron 
of K3 fiber of Calabi-Yau 3-fold. 
\newline
\noindent
$ \{ (x_3,x_4) \} 
\mid_{x_1=x_2=0} \in {}^2\nabla $~$ \subset {}^4\nabla~$  
 form  the integral points in  the  two-dimensional 
dual polyhedron of a  common 
 elliptic fiber in  Calabi-Yau 3-folds and in K3fiber.
We confirmed that most Calabi-Yau 3-folds and  
Calabi-Yau 4-folds in table 8  have the dual 
polyhedrons of   K3 and of 
elliptic curve which satisfy the above conditions.
 More precisely, 
for case (2), the dual polyhedron contains two  dual sub-polyhedra  
of T$^2$  and a dual polyhedron of K3 satisfying the above conditions.
They are CP$^2$(1,1,2)[4]  and CP$^2$(1,2,3)[6]. CP$^2$(1,2,3)[6]  
  comes from the elliptic fiber 
of the K3 fiber. 
If  the  above condition  for  having elliptic fibration  
 is sufficient,  then  we can conclude that  
 the two kinds of elliptic fibrations 
 coming from  both sides of the direct product of the 
Borcea-Voisin construction are possible for Calabi-Yau 3-folds. 
One comes  from the \T$^2$ side and the other comes from 
the elliptic fibration of K3.  
However, this will not be not sufficient condition  for  having 
elliptic fibration and needs farther  conditions to be  sufficent.   
\T$^2$ =\CP$^1$(1,2,3)[6] in 
Calabi-Yau 3-fold in case (2) does not satisfy   the  necessary condition 
of having elliptic fiberation about canonical bundles which is 
quoted in the reference of  \cite{vafa2} 
\footnote{We follow 
the notations of this reference 
\cite{vafa2}.}. 
\begin {equation}
K_{\rm CY3}= \pi^\ast ( K_B + \Sigma_i a_i E_i) + {\rm error~ terms}=0
\end{equation}
\noindent
The dual polyhedron  of table 8  leads to  $n_T =1=h^{1,1}(B)-1 $.  
This is not consistent with the above equation  when we suppose that  
T$^2$=CP$^1$(1,2,3)[6] fibered with E$_8$ type degeneration  
and B=F$_0$ based for   case (2).

The conclusion of the property of Calabi-Yau 3-folds by 
Borcea-Voisin methods is that  they 
do not have the elliptic fibration coming  from   
K3 fiber side. Even if    they have them  as fibers, then there may be no   
 sections. Therefore, they cannot be represented in the Weierstrass 
form as the extension of this elliptic curve with  the singularity. 
 The degenerations of  elliptic fibers  
of Calabi-Yau 3-folds do not relate to the degenerations of K3 fibers.    
 
For the dual polyhedron  of Calabi-Yau 4-folds of 
 in table 6 which constructed by the method in appendix B. 
It has two dual sub polyhedra of K3 
 as  two   fibers.    
Table 9 and Table 10 are a polyhedron and a dual polyhedron
of Calabi-Yau 3-fold (1) constructed by \K3=\CP$^3$(1,1,4,6)[12]
and \T$^2$=\CP$^2$(1,1,2)[12]. $\{ x_2,x_3\}$  and $ \{x_4 \}$ may 
relate  polyhedron and dual polyhedron of three tori.

\pagebreak


\begin{quote}
 
\begin{table}[h]
\[
\begin{array}
{|rrrr|rrrr|rrrr|rrrr|}
\hline  
\multicolumn{4}{|c|}{CP^2(1,2,3)[6]}                  
&\multicolumn{4}{|c|}{ CP^2(1,1,2)[4]}               
&\multicolumn{4}{|c|} {CP^3(1,6,14,21)[42]}              
&\multicolumn{4}{|c|}{CP^4(21,21,2,12,28)[84]} 
\\ \hline
  &  &  x_3 & x_4 
& x_1 &  &   & x_4  
&  & x_2 &  x_3 & x_4   
& x_1 & x_2 &  x_3 & x_4  
\\ \hline
 & & & 
&-1 & & &-2
& & & &
&-1 &0 &0 &-2
\\ \hline
 & & &
& & & &
& &-6 &-14 &-21
&0 &-6 &-14 &-21
\\ \hline
 & & & 
& & & &
& &-5 &-10 &-18
&0 &-5 &-12 &-18
\\ \hline
 & & & 
& & & &
& &-4 &-10 &-15
&0 &-4 &-10 &-15
\\ \hline
 & & & 
& & & &
& &-4 &-9 &-14
&0 &-4 &-9 &-14
\\ \hline
 & & & 
& & & &
& &-3 &-8 &-12
&0 &-3 &-8 &-12
\\ \hline
 & & & 
& & & &
& &-3 &-7 &-11
&0 &-3 &-7 &-11
\\ \hline
 & & & 
& & & &
& &-2 &-6 &-9
&0 &-2 &-6 &-9
\\ \hline
 & & & 
& & & &
& &-2 &-5 &-8
&0 &-2 &-5 &-8
\\ \hline
 & & & 
& & & &
& &-2 &-5 &-7
&0 &-2 &-5 &-7
\\ \hline
 & & & 
& & & &
& &-2 &-4 &-7
&0 &-2 &-4 &-7
\\ \hline
 & & & 
& & & &
& &-1 &-4 &-6
&0 &-1 &-4 &-6
\\ \hline
 & & & 
& & & &
& &-1 &-3 &-5
&0 &-1 &-3 &-5
\\ \hline
 & & & 
& & & &
& &-1 &-3 &-4
&0 &-1 &-3 &-4
\\ \hline
 & & & 
& & & &
& &-1 &-2 &-4
&0 &-1 &-2 &-4
\\ \hline
 & & & 
& & & &
& &-1 &-2 &-3
&0 &-1 &-2 &-3
\\ \hline
 & &-2 &-3 
& & & &
& &0 &-2 &-3
&0 &0 &-2 &-3
\\ \hline
 & &-1 &-2 
& & & &
& &0 &-1 &-2
&0 &0 &-1 &-2
\\ \hline
 & &-1 &-1 
& & & &
& &0 &-1 &-1
&0 &0 &-1 &-1
\\ \hline
 & &0 &-1 
&0 & & &-1
& &0 &0 &-1
&0 &0 &0 &-1
\\ \hline
 & &0 &0 
&0 & & &0
& &0 &0 &0
&0 &0 &0 &0
\\ \hline
 & &0 &1 
&0 & & &1
& &0 &0 &1
&0 &0&0 &1
\\ \hline
 & & 1& 0
& & & &
& &0 &1 &0
&0 &0 &1 &0
\\ \hline
 & &  & 
& & & &
& &1 &0 &0
&0 &1 &0 &0
\\ \hline
       &        &        &                    
& 1      &        &        &0 
& & & &
&  1   & 0       & 0       & 0                      
\\ \hline
\end{array}
\]
\caption{Dual polyhedron  
 of CY3-fold(2) and their sub dual polyhedra}
\end{table}

 
\begin{table}[h]
\[
\begin{array}
{|rrrr|rrrr|rrrr|rrrr|}
\hline  
\multicolumn{4}{|c|}{CP^2(1,2,3)[6]}                  
&\multicolumn{4}{|c|}{ CP^2(1,1,2)[4]}               
&\multicolumn{4}{|c|} {CP^3(1,1,4,6)[42]}              
&\multicolumn{4}{|c|}{{\rm CY 3-fold }(1)} 
\\ \hline
  &x_2  &  x_3 &  
&  & x_2 &   & x_4  
& x_1 & x_2 &  x_3 &    
& x_1 & x_2 &  x_3 & x_4  
\\ \hline
 &1 &-1 & 
& &1 &&-1
& -1&1 &-1 &
&-1 &1 &-1 &-1
\\ \hline
& -1&-1 &
& &-1 & &-1
& 0& -1&-1 &
&0 &-1 &-1 &-1
\\ \hline
 & -1&-1 & 
& &-1 & &0
& 0&-1 &-1 &
&0 &-1 &-1 &0
\\ \hline
 & -1&-1 & 
& &-1 & &1
& 0&-1 &-1 &
&0 &-1 &-1 &1
\\ \hline
&-1 &-1 & 
& &-1 & &2
&0 &-1 &-1 &
&0 &-1 &-1 &2
\\ \hline
 &-1 &-1 & 
& &-1 & &-3
&0 &-1 &-1 &
&0 &-1 &-1 &3
\\ \hline
 &0 &-1 & 
& &0 & &-1
&0 &0 &-1 &
&0 &0 &-1 &-1
\\ \hline
 &0 &-1 & 
& &0 & &0
&0 &0 &-1 &
&0 &0 &-1 &0
\\ \hline
 &0 &-1 & 
& &0 & &1
&0 &0 &-1 &
&0 &0 &-1 &1
\\ \hline
 &0 &0 & 
& &0 & &-1
&0 &0 &0 &
&0 &0 &0 &-1
\\ \hline
 &0 &0 & 
& &0 & &0
&0 &0 &0 &
&0 &0 &0 &0
\\ \hline
 &0 &0 & 
& &0 & &1
& 0&0 &0 &
&0 &0 &0 &1
\\ \hline
 &1 &-1 & 
& &1 & &-1
&0 &1 &-1 &
&0 &1 &-1 &-1
\\ \hline
 &1 &0 & 
& &1 & &-1
&0 &1 &0 &
&0 &1 &0 &-1
\\ \hline
 &1 &1 & 
& &1 & &-1
&0 &1 &1 &
&0 &1 &1 &-1
\\ \hline
 &1 &2 & 
& &1 & &-1
&0 &1 &2 &
&0 &1 &2 &-1
\\ \hline
 &1 &-1 & 
& &1 & &-1
&1 &1 &-1 &
&1 &1 &-1 &-1
\\ \hline
\end{array}
\]
\caption{a polyhedron  
 of CY3-fold(1)= $CP^1(1,1,2) \times CP^3(1,1,4,6)/ (-1) \times \sigma $
 and their sub  polyhedra}
\end{table}                 
\end{quote}

{\small 
\begin{quotation}

\begin{table}[h]
\[
\begin{array}
{|rrrr|rrrr|rrrr|rrrr|}
\hline  
\multicolumn{4}{|c|}{CP^2(1,2,3)[6]}                  
&\multicolumn{4}{|c|}{ CP^2(1,1,2)[4]}               
&\multicolumn{4}{|c|} {CP^3(5,6,22,33)[66]}              
&\multicolumn{4}{|c|}{{\rm CY 3-fold (1)}^\ast} 
\\ \hline
  &x_2  &  x_3 &  
&  &  & x_3  & x_4  
&x_1  & x_2 &  x_3 &    
& x_1 & x_2 &  x_3 & x_4  
\\ \hline
 & & & 
& & &&
& -6&3 &-2 &
&-6 &3 &-2 &0
\\ \hline
& & &
& & & &
& 5& 3&-2 &
&-5 &3 &-2 &0
\\ \hline
 & & & 
& & & &
& -4&2 &-1 &
&-4 &2 &-1 &0
\\ \hline
& & & 
& & & &
&-4 &3 &-2 &
&-4 &3 &-2 &0
\\ \hline
 & & & 
& & & &
&-3 &1 &-1 &
&-3 &1 &-1 &0
\\ \hline
 && & 
& & & &
&-3 &2 &-1 &
&-3 &2 &-1 &0
\\ \hline
 & & & 
& & & &
&-3 &3  &-2 &
&-3 &3 &-2 &0
\\ \hline
 & & & 
& & & &
&0 &0 &-1 &
&-2 &1 &-1 &0
\\ \hline
 & & & 
& & & &
&-2 &1 &0 &
&-2 &1 &0 &0
\\ \hline
 & & & 
& & & &
&-2 &2 &-1 &
&-2 &2 &-1 &0
\\ \hline
 & & & 
& & & &
& -2&3 &-2 &
&-2 &3 &-2 &0
\\ \hline
 & & & 
& & & &
&-1 &1 &-1 &
&-1 &1 &-1 &0
\\ \hline
 & & & 
& & & &
&-1&1 &-1 &
&-1 &1 &-1 &0
\\ \hline
 & & & 
& & & &
&-1 &1 &0 &
&-1 &1 &0 &0
\\ \hline
 & & & 
& & & &
&-1 &2 &-1 &
&-1 &2 &-1 &0
\\ \hline
 & &  & 
& & &  &
&1 &3 &-2 &
&-1 &3 &-2 &0
\\ \hline
  & & & 
& & -2  &  &-1
& & & &
&0 &-2 &0 &-1
\\ \hline
    & -1  &0 & 
&   & -1    &  &0
& 0 & -1  &0 &
&0  &-1   &0 &0
\\ \hline
 & 0&0 & 
& &0 & &0
& 0&0 &0 &
&0 &0 &0 &0
\\ \hline
& &  & 
& &0 &  &1
& & & &
&0 &0 &0 &1
\\ \hline
 & 0 &1 & 
& & &  &
&0 &0&1 &
&0 &0 &1 &0
\\ \hline
 &1 &-1 & 
& & & &
&0 &1 &-1 &
&0 &1 &-1 &0
\\ \hline
 &1 &0 & 
& &1 & &0
&0 &1 &0 &
&0&1 &0 &0
\\ \hline
 &2 &-1 & 
& & & &
&0 &2 &-1 &
&0&2 &-1 &0
\\ \hline
 &3 &-2 & 
& & & &
&0 &3 &-2 &
&0 &3 &-2 &0
\\ \hline
 & & & 
& & & &
&1 &0 &0 &
&1 &0&0 &0
\\ \hline
 & & & 
& & & &
&1 & 1& -1&
&1 &1 &-1 &0
\\ \hline
 & & & 
& & & &
&1 &1 &0 &
&1 &1 &0 &0
\\ \hline
   & & & 
&  & &   &
& 1 &2 &-1 &
&1 &2 &-1 &0
\\ \hline
 & & & 
& & & &
&1 &3 &-2 &
&1 &3 &-2 &0
\\ \hline
   & & & 
&  & &   & 
&2  &1 &-1 &
&2 &1 &-1 &0
\\ \hline
  & & & 
& & & &
&2 &1 &0 & 
&2 &1 &0 &0
\\ \hline
   & & & 
&  & &   &
&2  &2 &-1 &
&2 &2 &-1 &0
\\ \hline
&    &  &
&    & &   &  
& 2   & 3 &-2 &  
&2   &3  &-2 &  0
\\ \hline
 & & & 
& & & &
& 3&1 &-1 &
&3 &1 &-1 &0
\\ \hline
&   & & 
&   &   &  &
&3   &2  &-1 &
&3  &2  &-1 &0
\\ \hline
 & & & 
& & & &
&3 &3 &-2 &
&3 &3 &-2 &0
\\ \hline
 & & & 
& & & &
&4 &2 &-1 &
&4 &2 &-1 &0
\\ \hline
 & & & 
& & & &
&4 &3 &-2 &
&4 &3 &-2 &0
\\ \hline
 & & & 
& & & &
&5 &3 &-2 &
&5 &3 &-2 &0
\\ \hline
 & & & 
& & & &
&6 &3 &-2 &
&6 &3 &-2 &0
\\ \hline
\end{array}
\]
\caption{a dual polyhedron of CY3-fold(1) and their sub  dual polyhedra}
\end{table}

\end{quotation}
}

\end{document}